\documentclass[12pt]{article}%
\usepackage{graphics}%
\def\preprint%
#1{\thispagestyle{empty}~\newline\vspace*{-22.65mm}%
\begin{flushright}%
\begin{tabular}{l} #1%
\end{tabular}%
\end{flushright}%
\vspace{1cm}}%
\begin{document}%
\markright{\hfil Deriving relativistic momentum and energy}%
\title{\bf \LARGE Deriving relativistic momentum and energy}%
\author{Sebastiano Sonego\thanks{\tt
sebastiano.sonego@uniud.it}\hspace{2mm} %
and %
Massimo Pin\thanks{\tt pin@fisica.uniud.it}\\[2mm]%
{\small \it Universit\`a di Udine, Via delle Scienze 208,
33100 Udine, Italy}}%

\date{{\small June 24, 2004; \LaTeX-ed \today}}%
\maketitle%
\begin{abstract}%
We present a new derivation of the expressions for momentum and
energy of a relativistic particle.  In contrast to the procedures
commonly adopted in textbooks, the one suggested here requires
only the knowledge of the composition law for velocities along one
spatial dimension, and does not make use of the concept of
relativistic mass, or of the formalism of four-vectors.  The basic
ideas are very general and can be applied also to kinematics
different from the Newtonian and Einstein ones, in order to
construct the corresponding dynamics.\\%

\end{abstract}%
\def\p{\mbox{\boldmath $p$}}%
\def\u{\mbox{\boldmath $u$}}%
\def\v{\mbox{\boldmath $v$}}%
\def\w{\mbox{\boldmath $w$}}%
\def\n{\mbox{\boldmath $n$}}%
\def\F{\mbox{\boldmath $F$}}%
\def\bu{\bar{\mbox{\boldmath $u$}}}%
\def\0{\mbox{\boldmath $0$}}%
\def\Cdot{\mbox{\boldmath $\cdot$}}%
\def\Times{\mbox{\boldmath $\times$}}%
\def\Pphi{\mbox{\boldmath $\Phi$}}%
\def\d{{\mathrm d}}%
\def\SIZE{1.00}%

\section{Introduction}%
\label{sec:intro}%

Every physics student knows that, in Newtonian dynamics, a
particle with mass $m$ and velocity $\u$ has a momentum%
\begin{equation}%
\p=m\,\u%
\label{1}%
\end{equation}%
and a kinetic energy%
\begin{equation}%
T=\frac{1}{2}\,m\,u^2\;.%
\label{2}%
\end{equation}%
Usually, she does not stop to worry why these quantities are
defined just by Eqs.\ (\ref{1}) and (\ref{2}), and not by other
functions such as, e.g., $m\u/2$, $mu^2\u$, or $3mu^4$. This
comfortable situation ends when she attends a first course in
Einstein dynamics.  The expressions for momentum and kinetic
energy,%
\begin{equation}%
\p=\frac{m\,\u}{\sqrt{1-u^2/c^2}}%
\label{p}%
\end{equation}%
and%
\begin{equation}%
T=\frac{m\,c^2}{\sqrt{1-u^2/c^2}}-m\,c^2\;,%
\label{3}%
\end{equation}%
now look so unnatural that she wonders about the reasons for
choosing such complicated functions of velocity.  At this point
she can find, basically, three kinds of justifications for the
expressions (\ref{p}) and (\ref{3}) in textbooks dealing with
relativistic dynamics at an introductory level:%
\begin{enumerate}%
\item Requiring momentum conservation in all inertial frames for
an elastic collision where particles are deflected from their
initial line of flight \cite{feynman}.  Within this approach, the
discussion is not entirely trivial, because of the two-dimensional
character of the process.  Also, it seems odd that one need invoke
a second space dimension at all.  What happens if we restrict
ourselves to study motion along a straight line?  There ought to
be a way to find $p$ and $T$ without going ``outside'' in the
second dimension.%
\item Requiring momentum conservation for a head-on elastic
collision together with conservation of a ``relativistic mass''
\cite{bohm}.  This circumvents the previous difficulty, but the
use of a relativistic mass, and the pedagogical value of such a
concept, have been strongly criticised \cite{mass}.  (However, see
Ref.\ \cite{massdebate} for different opinions about this issue.)
Of course, what is used in this approach is actually conservation
of {\em energy\/} $E$ (equal to $T+m\,c^2$), but why then should
one assume that $\p=E\u/c^2$, as done implicitly by the authors in
Ref.\ \cite{bohm}?%
\item Working with four-vectors, so one defines four-momentum
just in the same way as the three-momentum of Newtonian theory,
but with three-velocity replaced by four-velocity \cite{ew}. The
problem here is that there is no guarantee, {\em a priori\/}, that
such a quantity will be conserved for an isolated system. Indeed,
the conservation law is usually checked for the case of a simple
elastic collision, {\em after\/} four-momentum has been defined.
Also, this approach requires the introduction of radically new
ideas, hence it is unsuitable for a conceptually
elementary presentation of the theory.%
\end{enumerate}%

There is a fourth approach, which to our knowledge has never been
adopted in textbooks,\footnote{Actually, the idea has not been
totally ignored; see, e.g., Ref.\ \cite{erp}.  However, its extant
implementations are not as simple as they could, making use of
two-dimensional scattering and four-vectors.} that resembles 1
above but is cleaner and can be consistently applied even in one
space dimension only.  It is based on the remark that, if energy
is conserved in {\em all\/} inertial frames, then ``something
else'' is also conserved. In the non-relativistic regime, this
``something else'' turns out to coincide with linear momentum.  We
suggest doing the same at the relativistic level.%

We believe that this is the best strategy to adopt in an
introductory course, because it relies on the same physical
concepts that students are already familiar with from Newtonian
mechanics, and does not introduce any new notion like relativistic
mass or four-vectors. It focusses on similarities, rather than
differences, between Newtonian and Einstein dynamics. Hence, a
student is not required to replace physical pictures and
mathematical tools with other, quite different ones --- minor
adaptations are enough, the general scheme remaining the same.%

Furthermore, the approach we suggest casts light not only on the
expressions (\ref{p}) and (\ref{3}) for momentum and energy in
special relativity, but on (\ref{1}) and (\ref{2}) in Newtonian
mechanics as well.  Indeed, the underlying philosophy is that
energy and momentum are nothing else than functions of mass and
velocity that, under suitable conditions, happen to be conserved.
This is why we treat in a special way those functions, rather than
others. This point of view deserves to be emphasised in a
pedagogical exposition, because it provides clear insights on the
reasons why momentum and energy are defined the way they are, at
the same time demystifying their meaning.%

Our starting point is the definition of kinetic energy for a
particle, as a scalar quantity whose change equals the work done
on the particle.  Mathematically, one first defines the power
(work per unit time)%
\begin{equation}%
W:=\F\,\Cdot\,\u\;,%
\label{W}%
\end{equation}%
where $\F$ is the total force acting on the particle.  Then, using
Newton's second law%
\begin{equation}%
\frac{{\rm d}\p}{{\rm d}t}=\F\;,%
\label{Newton}%
\end{equation}%
which holds both in the non-relativistic and in the relativistic
regimes, one gets also%
\begin{equation}%
W=\frac{{\rm d}{\p}}{{\rm d}t}\,\Cdot\,\u\;.%
\label{pu}%
\end{equation}%
Finally, one defines a function $T(\u)$ such that%
\begin{equation}%
\frac{{\rm d}T}{{\rm d}t}:=W%
\label{forzevive}%
\end{equation}%
(this is possible since using Newton's second law we have equated
power to a purely kinetic quantity), thus obtaining%
\begin{equation}%
\d T=\u\,\Cdot\,\d\p\;.%
\label{dT}%
\end{equation}%
In the following, we shall adopt Eq.\ (\ref{dT}) as a fundamental
relationship between kinetic energy and momentum, that stands up
on its own, independently of any justification like the
one based on Eqs.\ (\ref{W})--(\ref{forzevive}).%

For a system of non-interacting particles, kinetic energy is
necessarily additive, since work is.  We are then ready to
introduce the two physical postulates upon which our
discussion is based:%
\begin{description}%
\item[{\rm P1.}] The principle of relativity;%
\item[{\rm P2.}] The existence of elastic collisions between
asymptotically free particles.%
\end{description}%
As we shall see, from these ingredients it follows that in an
elastic collision, there is another quantity that is conserved ---
a vector one, that we identify with momentum.  This quantity is
linked to kinetic energy through a simple equation containing a
function\footnote{In the one-dimensional case discussed in this
paper.  In three dimensions, $\varphi$ is replaced by a matrix of
functions $\varphi_{ij}$, where the indices $i$ and $j$ run from 1
to 3.} $\varphi$, that follows from the composition law for
velocities. This, together with Eq.\ (\ref{dT}) above, allows us
to find the explicit dependence of both momentum and energy on
velocity.  We can then easily construct also the expressions for
the Lagrangian and the Hamiltonian for a free particle in an
inertial frame. Hence, the entire basis of dynamics turns out to
be uniquely determined by the function $\varphi$, a purely
kinematical quantity.%

We stress again that this procedure works for Newtonian and
Einstein mechanics as well, so a student who is already familiar
with Newtonian concepts will have no difficulty in following the
argument, since no new notion is required.  The treatment has
therefore a unifying conceptual power.  In addition, it shows
clearly and explicitly why Newtonian and Einstein dynamics are
different.  This happens because the expression we derive for
momentum depends on the composition law for velocities (through
the function $\varphi$), which is not the same in Galilean and
Lorentz kinematics. Hence, Newtonian dynamics is not the same as
Einstein dynamics only because the underlying kinematics are
different.  In fact, {\em all\/} differences between the two
theories can be traced back to this one --- another point well
worth emphasising when teaching the subject.  Of course, this is
just an alternative way for saying that one can construct a
relativistic theory based on Galilei invariance, or on Lorentz
invariance.  However, a beginner will find it easier to follow
this approach, rather than more abstract considerations of formal
symmetry.%

The generality of the method makes it, in fact, applicable to a
wider class of theories.  Indeed, as discussed by Mermin
\cite{mermin}, the principle of relativity is compatible with a
generalised kinematics, that includes the Galilean and Lorentz
ones as particular cases.  One can then apply the techniques
discussed in the present paper, in order to construct the
corresponding dynamics.  This will be done systematically in Ref.\
\cite{jmp}.  Here, we focus only on the Newtonian and Einstein
cases, which are by far the most important from a pedagogical
point of view.%

The paper is structured as follows.  In Sec.\ \ref{sec:newtonian}
we present the basic ideas.  In Sec.\ \ref{sec:comp-law} we review
the main points behind Mermin's discussion of the composition law
for velocities.  We do this both for the sake of completeness and
for notational convenience, because the function $\varphi$ that
plays an essential role in our method also emerges naturally in
Mermin's approach to kinematics.  However, it should be obvious to
the reader that, since the core of the paper depends only on the
formula for the composition law, an instructor who wants to follow
our suggestions in the particular cases of Newtonian and Einstein
mechanics does not necessarily have to present kinematics as in
Ref.\ \cite{mermin}, and can adopt a more traditional approach. In
Sec.\ \ref{sec:general} we present the general derivation of the
expressions for momentum, kinetic energy, the Lagrangian, and the
Hamiltonian.  Then, in Sec.\ \ref{sec:examples} we apply these
results to construct the basis of Newtonian and Einstein dynamics.
Again, an instructor can easily shorten significantly the
presentation of this material at his/her convenience, according to
the level of sophistication of the class.  Section
\ref{sec:comments} contains some concluding comments about the
different status of energy and momentum conservation, and the fact
that there are no other conservation laws in one spatial
dimension.%

With the exception of Sec.\ \ref{sec:newtonian}, we restrict
ourselves to considering motion along one space dimension. This
makes the material accessible to a student with an elementary
knowledge of calculus. In particular, no knowledge of vector
algebra is required, contrary to what happens in approaches 1 and
3 above.  The extension to three space dimensions is almost
straightforward, but we prefer to postpone it to another
publication for pedagogical clarity.%

\section{Main ideas}%
\label{sec:newtonian}%
\setcounter{equation}{0}%

Consider the following argument in Newtonian mechanics, originally
due to Huygens \cite{barbour}.  Conservation of energy in an
inertial frame $\cal K$ during an elastic collision between
two particles with masses $m_1$ and $m_2$ gives%
\begin{equation}%
\frac{1}{2}\,m_1 \u_1^2+\frac{1}{2}\,m_2 \u_2^2=\frac{1}{2}\,m_1
{\u'}_1^2+\frac{1}{2}\,m_2 {\u'}_2^2\;.%
\label{Tnewt}%
\end{equation}%
With respect to another inertial frame $\overline{\cal K}$, in
which $\cal K$ moves with velocity $\v$, the velocities are
$\bar{\u}_1=\u_1+\v$, $\bar{\u}_2=\u_2+\v$,
$\bar{\u}'_1={\u'}_1+\v$, and $\bar{\u}'_2={\u'}_2+\v$.
Conservation of energy in $\overline{\cal K}$ then implies%
\begin{equation}%
\frac{1}{2}\,m_1\left(\u_1+\v\right)^2+\frac{1}{2}\,m_2
\left(\u_2+\v\right)^2=\frac{1}{2}\,m_1\left(\u'_1+\v\right)^2
+\frac{1}{2}\,m_2\left(\u'_2+\v\right)^2\;.%
\label{T'newt}%
\end{equation}%
Expanding the squares and using Eq.\ (\ref{Tnewt}), one
immediately gets%
\begin{equation}%
\left(m_1 \u_1+m_2 \u_2\right)\Cdot\,\v
=\left(m_1 \u'_1+m_2 \u'_2\right)\Cdot\,\v\;.%
\label{derived}%
\end{equation}%
Since this must hold for an arbitrary vector $\v$, momentum
conservation follows immediately.%

The idea behind this derivation of the expression for momentum can
easily be generalised to any theory satisfying postulates P1 and
P2.  More precisely, let $T(\u)$ be the kinetic energy of a
particle with velocity $\u$ in an inertial frame $\cal K$. In an
elastic collision,%
\begin{equation}%
T_1({\u}_1)+T_2({\u}_2)=T_1({\u}'_1)+T_2({\u}'_2)\;.%
\label{T+T}%
\end{equation}%
(Of course, the kinetic energy will also depend on the particle
mass; we keep track of this dependence with the indices 1 and 2 on
$T$.)  With respect to $\overline{\cal K}$,%
\begin{equation}%
T_1(\bar{\u}_1)+T_2(\bar{\u}_2)=T_1(\bar{\u}'_1)+T_2(\bar{\u}'_2)\;,%
\label{T'+T'}%
\end{equation}%
where now $\bar{\u}=\Pphi(\u,\v)$ is the composition law for
velocities.  On expanding Eq.\ (\ref{T'+T'}) in the variable $\v$
and using Eq.\ (\ref{T+T}), we find a conservation equation to
first order in $\v$, analogous to the one expressed by Eq.\
(\ref{derived}) --- although with different coefficients, in
general.  We can then define linear momentum\footnote{With this
definition, linear momentum turns out to be a one-form rather than
a vector, which is very satisfactory from a formal point of view.}
$\p$ as the first-order coefficient in $\v$. If we know the
function $T(\u)$, we can find $\p$.%

If we do not already know $T(\u)$, we can define it by requiring
that it obey the fundamental relation (\ref{dT}).  Then, putting
together Eq.\ (\ref{dT}) and the one that expresses $\p$ in terms
of $T(\u)$, one obtains a simple system of differential equations,
from which it is possible to find both $T(\u)$ and $\p(\u)$.  In
the following sections, we implement these ideas in detail,
restricting ourselves to considering motion along one spatial
dimension.%

\section{Velocity composition law}%
\label{sec:comp-law}%
\setcounter{equation}{0}

Suppose that a particle moves with velocity $u$ with respect to a
reference frame $\cal K$.  If $\cal K$ moves with velocity $v$
with respect to another reference frame $\overline{\cal K}$, the
particle velocity $\bar{u}$ with respect
to $\overline{\cal K}$ is given by some composition law%
\begin{equation}%
\bar{u}=\Phi(u,v)\;.%
\label{comp-gen}%
\end{equation}%

Of course, when $\cal K$ is at rest with respect to
$\overline{\cal K}$, we have that $\bar{u}=u$.  Similarly, if the
particle is at rest in $\cal K$, its velocity with respect to
$\overline{\cal K}$ is the same of $\cal K$.  More synthetically,%
\begin{equation}%
\Phi(u,0)=\Phi(0,u)=u\;,\qquad\forall u\;.%
\label{id}%
\end{equation}%

Also, a particle at rest in $\overline{\cal K}$ has $\bar{u}=0$,
hence it moves with respect to $\cal K$ with a velocity $v_L$ such
that%
\begin{equation}%
\Phi(v_L,v)=0\;,\qquad\forall v\;.%
\label{L-inverse}%
\end{equation}%
Similarly, if a particle moves with respect to $\cal K$ with
velocity $u$ but is at rest in $\overline{\cal K}$, then the speed
$u_R$ of $\cal K$ with respect to $\overline{\cal K}$ must satisfy
the relation%
\begin{equation}%
\Phi(u,u_R)=0\;,\qquad\forall u\;.%
\label{R-inverse}%
\end{equation}%

Finally, from the relativity principle it follows \cite{mermin}
that $\Phi$ satisfies the associative law, i.e., that%
\begin{equation}%
\Phi(\Phi(u,v),w)=\Phi(u,\Phi(v,w))\;,\qquad\forall u,v,w\;.%
\label{ass}%
\end{equation}%
Therefore, Eq.\ (\ref{comp-gen}) gives the composition law of a
group, with neutral element $0$ and left- and right-inverses given
by Eqs.\ (\ref{L-inverse}) and (\ref{R-inverse}).  Combining Eqs.\
(\ref{L-inverse})--(\ref{ass}) one finds that there actually is
only one inverse of $v$, say $v'$ --- that is, $v_L=v_R=:v'$.
Note that the inverse $v'$ of $v$ is not necessarily equal to
$-v$.%

Let us now define the function%
\begin{equation}%
\varphi(u):=\left.\frac{\partial \Phi(u,v)}{\partial
v}\right|_{v=0}\;.%
\label{f(u)}%
\end{equation}%
The meaning of $\varphi$ can be found by expanding $\bar{u}$ to
the first order in $v$:%
\begin{equation}%
\bar{u}=u+\varphi(u)\,v+{\cal O}(v^2)\;.%
\label{expu}%
\end{equation}%
This is the composition law between an {\em arbitrary\/} velocity
$u$ and a velocity $v$ with {\em small\/} magnitude.  Since Eq.\
(\ref{id}) implies $\varphi(0)=1$, at very low speeds one
recovers Galilean kinematics.%

Another interesting property of $\varphi$ is that if some
velocity, say $C$, is invariant, then $\varphi(C)=0$. This follows
immediately by applying Eq.\ (\ref{f(u)}) to the condition%
\begin{equation}%
\Phi(C,v)=C\;,\qquad\forall v\;,%
\label{fixed}%
\end{equation}%
which expresses the invariance of $C$.%

The function $\varphi$ contains all the information needed to
specify $\Phi$.  Indeed, it is not difficult to show \cite{mermin}
that $\Phi$ can be written as%
\begin{equation}%
\Phi(u,v)=h^{-1}\left(h(u)+h(v)\right)\;,%
\label{hh}%
\end{equation}%
where%
\begin{equation}%
\frac{\d h(u)}{\d u}=\frac{1}{\varphi(u)}%
\label{h'}%
\end{equation}%
and $h(0)=0$.  As a corollary of Eq.\ (\ref{hh}), one finds that
the composition law for collinear velocities is also commutative,
i.e.,%
\begin{equation}%
\Phi(u,v)=\Phi(v,u)\;,\qquad\forall u,v\;.%
\label{comm}%
\end{equation}%
It is worth stressing, however, that this property does not hold
in general for the composition law of velocities along arbitrary
directions in more than one spatial dimension \cite{rindler}.%

With further requirements, essentially equivalent to homogeneity
of space and time, and spatial isotropy (or better, its
one-dimensional counterpart --- the physical equivalence of the
two orientations in the one-dimensional space), one can further
restrict $\varphi$ to the form%
\begin{equation}%
\varphi(u)=1-K\,u^2\;,%
\label{K}%
\end{equation}%
where $K$ is a constant \cite{mermin}.  Moreover, the
possibility $K<0$ can be excluded on physical grounds.%

The case $K=0$ gives the simple Galilean addition of velocities.
With $K>0$ one finds%
\begin{equation}%
h(u)=\frac{1}{\sqrt{K}}\ln\left(\frac{1+\sqrt{K}\,u}{1-
\sqrt{K}\,u}\right)^{1/2}\;.%
\label{U}%
\end{equation}%
This leads to Einstein's composition law, with the speed of light
replaced by $1/\sqrt{K}$.  Hence, the mathematical structure of
Einstein's composition law is a consequence of the principle of
relativity alone, combined with the postulates of homogeneity of
space and time, and of spatial isotropy.  Remarkably, this was
known to Kaluza already in 1924 \cite{kaluza}. An analogous result
about the structure of the Lorentz transformation was obtained by
von Ignatowsky in 1910 \cite{ignat}, and has been rediscovered
many times since \cite{rediscoverer}.  (See also Ref.\
\cite{gorini} for a rigorous derivation, and Ref.\
\cite{textbooks} for clear presentations at the textbook level.)%

\section{General analysis}%
\label{sec:general}%
\setcounter{equation}{0}

We now carry on the programme outlined in Sec.\
\ref{sec:newtonian}, deriving the general expressions for momentum
and kinetic energy (Sec.\ \ref{subsec:pT}), the Lagrangian (Sec.\
\ref{subsec:lagrangian}), and the Hamiltonian (Sec.\
\ref{subsec:hamiltonian}) for a free particle in an inertial
frame, that follow from postulates P1 and P2 when a given
composition law for velocities is adopted.%

\subsection{Momentum and kinetic energy}%
\label{subsec:pT}%

Let $T(u)$ be the kinetic energy of a particle in a reference
frame where the particle velocity is $u$. (Of course, $T(u)$ may
depend on some invariant parameters characterising the particle,
in addition to its velocity.  For example, in Newtonian dynamics
it depends on the particle mass.)  In an inertial frame, $T(u)$ is
conserved for a free particle, because $u$ is constant, by the
principle of inertia.  Then, the total kinetic energy is
conserved also for a system of non-interacting particles.%

According to postulate P2, there are spatially localised
interactions between particles which do not change the total
kinetic energy.  It is then easy to see that there is another
additive quantity which is conserved in any theory in which a
relativity principle holds (postulate P1).  More specifically, if
the composition law for velocities is given by Eq.\
(\ref{comp-gen}), such a quantity is, for a single particle,%
\begin{equation}%
p(u)=\varphi(u)\,\frac{\d T(u)}{\d u}\;,%
\label{momentum}%
\end{equation}%
where $\varphi(u)$ is the function defined by Eq.\ (\ref{f(u)}).%

The proof of this theorem relies on the generalisation of Huygens'
argument outlined in Sec.\ \ref{sec:newtonian}.  Let us write
energy conservation in two inertial frames $\cal K$ and
$\overline{\cal K}$ during a head-on elastic collision between
two particles, as in Eqs.\ (\ref{T+T}) and (\ref{T'+T'}):%
\begin{equation}%
T_1(u_1)+T_2(u_2)=T_1(u'_1)+T_2(u'_2)\;;%
\label{EK}%
\end{equation}%
\begin{equation}%
T_1(\bar{u}_1)+T_2(\bar{u}_2)=T_1(\bar{u}'_1)+T_2(\bar{u}'_2)\;.%
\label{EH}%
\end{equation}%
We have used the same functions $T_1$ and $T_2$ in both reference
frames because of the relativity principle.  We can expand the
generic function $T(\bar{u})$ around $v=0$, and use the property
(\ref{id}) to get%
\begin{equation}%
T(\bar{u})=T(u)+\frac{\d T(u)}{\d u}\,\varphi(u)\,v+{\cal O}(v^2)\;.%
\label{exp}%
\end{equation}%
Doing this for each term in Eq.\ (\ref{EH}) and using Eq.\ (\ref{EK}),
then dividing by $v$ and taking the limit for $v\to 0$, we obtain%
\begin{equation}%
\frac{\d T_1(u_1)}{\d u_1}\,\varphi(u_1)+\frac{\d T_2(u_2)}{\d
u_2}\,\varphi(u_2)=\frac{\d T_1(u'_1)}{\d
u'_1}\,\varphi(u'_1)+\frac{\d
T_2(u'_2)}{\d u'_2}\,\varphi(u'_2)\;.%
\label{consmom}%
\end{equation}%
This proves the claim above.\footnote{Actually, this argument only
shows that the quantity on the right-hand side of Eq.\
(\ref{momentum}) is conserved.  The most general form of momentum
is then%
\[ p=\lambda\,\varphi\,\frac{\d T}{\d u}+\mu\,T+\nu\;,\]%
where $\lambda$, $\mu$, and $\nu$ are constants.  It turns out
that the simplest non-trivial choice $\lambda=1$, $\mu=\nu=0$ is
the one that leads to Newtonian and Einstein dynamics, while other
values of the coefficients correspond to alternative relativistic
dynamics (in particular, for $\mu\neq 0$ one finds anisotropic
theories). We leave the discussion of these possibilities to a
forthcoming, more technical paper \cite{jmp}.}%

Of course, Eq.\ (\ref{momentum}) is not sufficient in order to
find an expression for momentum, since the function $T(u)$ is also
unknown.  However, as already discussed in Secs.\ \ref{sec:intro}
and \ref{sec:newtonian}, we can define kinetic energy so that its
variation gives the work done on the particle --- that is, impose
the validity of Eq.\ (\ref{dT}). Combining Eq.\ (\ref{dT}) in its
one-dimensional version ($\d T=u\,\d p$) with Eq.\
(\ref{momentum}), we obtain a differential equation for the
function $p(u)$:%
\begin{equation}%
\frac{\d p}{\d u}=\frac{p}{u\,\varphi(u)}\;.%
\label{dp/du}%
\end{equation}%
Integrating by separation of variables, one finds the expression
for $p(u)$. In general, we can write%
\begin{equation}%
p(u)=m\,\exp\int^u \d u'\,\frac{1}{u'\,\varphi(u')}\;,%
\label{p=mf}%
\end{equation}%
where $m$ is a constant parameter that can vary from particle to
particle.  Mathematically, $m$ represents the arbitrary constant
associated with the general solution of the differential equation
(\ref{dp/du}).  Physically, it is identified with the particle
mass by imposing the Newtonian limit for $u\to 0$.%

Finally, one can replace the expression for $p(u)$ into Eq.\
(\ref{momentum}), and integrate with the condition $T(0)=0$ to
obtain also the expression for $T(u)$:%
\begin{equation}%
T(u)=\int_0^u\d u'\,\frac{p(u')}{\varphi(u')}\;.%
\label{T-gen}%
\end{equation}%
%

\subsection{Lagrangian}%
\label{subsec:lagrangian}%

The Lagrangian should satisfy the relation%
\begin{equation}%
p(u)=\frac{\d L(u)}{\d u}\;.%
\label{p=dL/du}%
\end{equation}%
Using Eq.\ (\ref{momentum}), we obtain%
\begin{equation}%
\d L(u)=\varphi(u)\,\frac{\d T(u)}{\d u}\,\d
u\;.%
\label{dL}%
\end{equation}%
Obviously, it is only for $\varphi=1$ that $L=T+\mbox{const}$, so
the Lagrangian for a free particle coincides with the kinetic
energy only in Newtonian dynamics.%

\subsection{Hamiltonian}%
\label{subsec:hamiltonian}%

Equation (\ref{dT}) allows us to identify the Hamiltonian for
a free particle.  Indeed,%
\begin{equation}%
u=\left.\frac{\d T(u)}{\d u}\right/\frac{\d p(u)}{\d u}=\frac{\d
T(u)}{\d u}\frac{\d u(p)}{\d p}=\frac{\d T(u(p))}{\d p}\;.%
\label{hamilton}%
\end{equation}%
On the other hand, one of Hamilton's equations of motion is%
\begin{equation}%
u=\frac{\d H(p)}{\d p}\;,%
\label{true-ham}%
\end{equation}%
so one can write $H(p)$ as $T(u(p))$, up to a $u$-independent
additive term.  Apart from $u$, the only other parameter $T$
depends on is the particle mass $m$, so we have in general%
\begin{equation}%
H(p,m)=T(u(p,m),m)+E_0(m)\;,%
\label{E_0}%
\end{equation}%
where we have made explicit the dependence of the various
quantities on $m$, and $E_0(m)$ denotes the value of the
Hamiltonian when $p=0$. Since, numerically, $H$ coincides with the
particle energy $E$, it follows from Eq.\ (\ref{E_0}) that%
\begin{equation}%
E(u,m)=T(u,m)+E_0(m)\;,%
\label{E}%
\end{equation}%
so $E_0(m)$ can be interpreted as the particle {\em rest energy\/}.%

Of course, the same expression for $H$ can be obtained as the
Legendre transform \cite{goldstein} of $L$.%

\section{Special cases}%
\label{sec:examples}%
\setcounter{equation}{0}

Let us apply the general results derived in the previous section
to the two cases of pedagogical interest, namely Newtonian
and Einstein dynamics.%

\subsection{Newtonian dynamics}%
\label{subsec:nonrelativistic-1d}%

The composition law is simply%
\begin{equation}%
\Phi(u,v)=u+v\;,%
\label{u+v}%
\end{equation}%
so $\varphi(u)=1$.  From Eq.\ (\ref{p=mf}) one then finds
immediately $p(u)=m\,u$ which, replaced into Eq.\ (\ref{T-gen}),
gives $T(u)=m\,u^2/2$.%

The Lagrangian coincides with the kinetic energy, as already
noted.  In order to get the Hamiltonian, we first express
velocity as a function of momentum, $u(p)=p/m$, so%
\begin{equation}%
H(p)=\frac{p^2}{2\,m}+E_0(m)\;.%
\label{HG}%
\end{equation}%
The choice $E_0(m)=0$ is obviously the simplest.  Note that
particles of zero mass cannot exist in this theory.%

\subsection{Einstein dynamics}%
\label{subsec:relativistic-1d}%

Einstein's composition law%
\begin{equation}%
\Phi(u,v)=\frac{u+v}{1+uv/c^2}%
\label{composition}%
\end{equation}%
corresponds to%
\begin{equation}%
\varphi(u)=1-u^2/c^2\;.%
\label{rel-f}%
\end{equation}%
(Note that $\varphi(\pm c)=0$, so $\pm c$ are invariant
velocities.)  From Eq.\ (\ref{p=mf}) one finds%
\begin{equation}%
p(u)=m\,u\,\gamma(u)\;,%
\label{momentum!}%
\end{equation}%
where we have defined the Lorentz factor%
\begin{equation}%
\gamma(u):=\left(1-u^2/c^2\right)^{-1/2}\;.%
\label{gamma}%
\end{equation}%
The expression for the kinetic energy follows immediately on
replacing Eq.\ (\ref{momentum!}) into Eq.\ (\ref{T-gen}):%
\begin{equation}%
T(u)=m\,c^2\,\gamma(u)-m\,c^2\;.%
\label{T(u)}%
\end{equation}%

The Lagrangian is%
\begin{equation}%
L(u)=-m\,c^2/\gamma(u)\;.%
\label{L-rel}%
\end{equation}%
For zero-mass particles, the function $L$ is ill-defined, and
a Lagrangian formulation is not viable.%

Inverting Eq.\ (\ref{momentum!}) we get
\begin{equation}%
u(p)=\frac{p\,c}{\sqrt{p^2+m^2 c^2}}\;,%
\label{uR}%
\end{equation}%
so the Hamiltonian is%
\begin{equation}%
H(p)=\sqrt{p^2 c^2+m^2 c^4}-mc^2+E_0(m)\;.%
\label{HR}%
\end{equation}%
Now, the simplest choice is $E_0(m)=mc^2$.  In this theory we can
treat also zero-mass particles, for which $H(p)=p\,c$.%

\section{Comments}%
\label{sec:comments}%
\setcounter{equation}{0}%

The most important result presented in this paper is the theorem
in Sec.\ \ref{sec:general} --- that, in theories obeying
postulates P1 and P2, the quantity $p(u)$ defined by Eq.\
(\ref{momentum}) is conserved and can be identified with momentum.
Broadly, this implies that, in such theories, kinematics
``determines'' dynamics.  As applications, we have shown how to
recover the expressions for the quantities on which dynamics is
based, in the two cases of Newtonian and Einstein mechanics.  Of
course, one may consider other types of dynamics as well
\cite{jmp}, based on alternative kinematics but still obeying
postulates P1 and P2.%

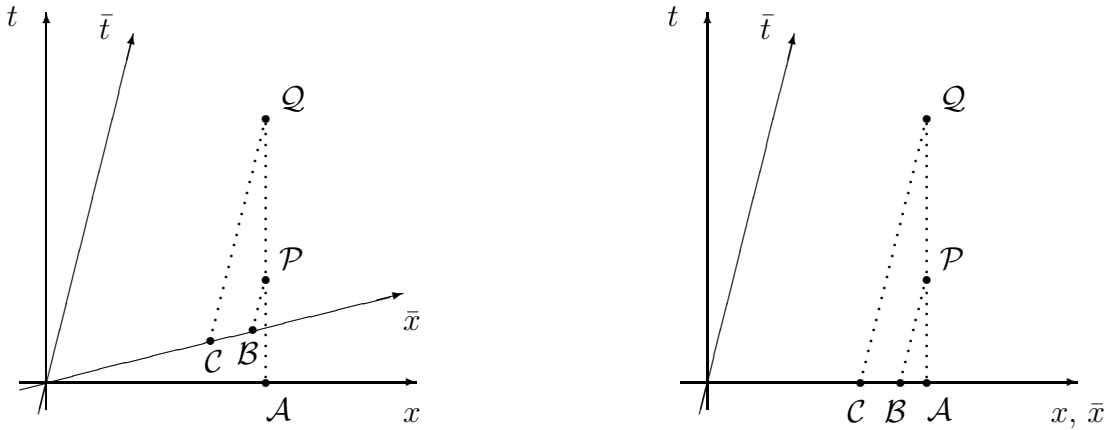
\begin{figure}%
\begin{picture}(150,200)(90,90)%
\put(110,120){\vector(1,0){150}}%
\put(255,105){$x$}%
\put(120,110){\vector(0,1){150}}%
\put(105,255){$t$}%
\put(120,120){\vector(4,1){135}}%
\put(120,120){\line(-4,-1){10}}%
\put(255,140){$\bar x$}%
\put(120,120){\vector(1,4){33}}%
\put(120,120){\line(-1,-4){3}}%
\put(140,250){$\bar t$}%
\put(203,159){\circle*{3}}%
\put(208,163){$\cal P$}%
\put(203,220){\circle*{3}}%
\put(208,224){$\cal Q$}%
\put(203,218){\circle*{1}}%
\put(203,214){\circle*{1}}%
\put(203,210){\circle*{1}}%
\put(203,206){\circle*{1}}%
\put(203,202){\circle*{1}}%
\put(203,198){\circle*{1}}%
\put(203,194){\circle*{1}}%
\put(203,190){\circle*{1}}%
\put(203,186){\circle*{1}}%
\put(203,182){\circle*{1}}%
\put(203,178){\circle*{1}}%
\put(203,174){\circle*{1}}%
\put(203,170){\circle*{1}}%
\put(203,166){\circle*{1}}%
\put(203,162){\circle*{1}}%
\put(203,158){\circle*{1}}%
\put(203,154){\circle*{1}}%
\put(203,150){\circle*{1}}%
\put(203,146){\circle*{1}}%
\put(203,142){\circle*{1}}%
\put(203,138){\circle*{1}}%
\put(203,134){\circle*{1}}%
\put(203,130){\circle*{1}}%
\put(203,126){\circle*{1}}%
\put(203,122){\circle*{1}}%
\put(203,120){\circle*{3}}%
\put(203,106){$\cal A$}%
\put(202,216){\circle*{1}}%
\put(201,212){\circle*{1}}%
\put(200,208){\circle*{1}}%
\put(199,204){\circle*{1}}%
\put(198,200){\circle*{1}}%
\put(197,196){\circle*{1}}%
\put(196,192){\circle*{1}}%
\put(195,188){\circle*{1}}%
\put(194,184){\circle*{1}}%
\put(193,180){\circle*{1}}%
\put(192,176){\circle*{1}}%
\put(191,172){\circle*{1}}%
\put(190,168){\circle*{1}}%
\put(189,164){\circle*{1}}%
\put(188,160){\circle*{1}}%
\put(187,156){\circle*{1}}%
\put(186,152){\circle*{1}}%
\put(185,148){\circle*{1}}%
\put(184,144){\circle*{1}}%
\put(183,140){\circle*{1}}%
\put(182,136){\circle*{3}}%
\put(180,124){$\cal C$}%
\put(202,156){\circle*{1}}%
\put(201,152){\circle*{1}}%
\put(200,148){\circle*{1}}%
\put(199,144){\circle*{1}}%
\put(198,140){\circle*{3}}%
\put(192,127){$\cal B$}%
\put(360,120){\vector(1,0){150}}%
\put(500,105){$x$, $\bar x$}%
\put(370,110){\vector(0,1){150}}%
\put(355,255){$t$}%
\put(370,120){\vector(1,4){33}}%
\put(370,120){\line(-1,-4){3}}%
\put(390,250){$\bar t$}%
\put(453,159){\circle*{3}}%
\put(458,163){$\cal P$}%
\put(453,220){\circle*{3}}%
\put(458,224){$\cal Q$}%
\put(453,218){\circle*{1}}%
\put(453,214){\circle*{1}}%
\put(453,210){\circle*{1}}%
\put(453,206){\circle*{1}}%
\put(453,202){\circle*{1}}%
\put(453,198){\circle*{1}}%
\put(453,194){\circle*{1}}%
\put(453,190){\circle*{1}}%
\put(453,186){\circle*{1}}%
\put(453,182){\circle*{1}}%
\put(453,178){\circle*{1}}%
\put(453,174){\circle*{1}}%
\put(453,170){\circle*{1}}%
\put(453,166){\circle*{1}}%
\put(453,162){\circle*{1}}%
\put(453,158){\circle*{1}}%
\put(453,154){\circle*{1}}%
\put(453,150){\circle*{1}}%
\put(453,146){\circle*{1}}%
\put(453,142){\circle*{1}}%
\put(453,138){\circle*{1}}%
\put(453,134){\circle*{1}}%
\put(453,130){\circle*{1}}%
\put(453,126){\circle*{1}}%
\put(453,122){\circle*{1}}%
\put(453,120){\circle*{3}}%
\put(453,106){$\cal A$}%
\put(452,216){\circle*{1}}%
\put(451,212){\circle*{1}}%
\put(450,208){\circle*{1}}%
\put(449,204){\circle*{1}}%
\put(448,200){\circle*{1}}%
\put(447,196){\circle*{1}}%
\put(446,192){\circle*{1}}%
\put(445,188){\circle*{1}}%
\put(444,184){\circle*{1}}%
\put(443,180){\circle*{1}}%
\put(442,176){\circle*{1}}%
\put(441,172){\circle*{1}}%
\put(440,168){\circle*{1}}%
\put(439,164){\circle*{1}}%
\put(438,160){\circle*{1}}%
\put(437,156){\circle*{1}}%
\put(436,152){\circle*{1}}%
\put(435,148){\circle*{1}}%
\put(434,144){\circle*{1}}%
\put(433,140){\circle*{1}}%
\put(432,136){\circle*{1}}%
\put(431,132){\circle*{1}}%
\put(430,128){\circle*{1}}%
\put(429,124){\circle*{1}}%
\put(428,120){\circle*{3}}%
\put(423,105){$\cal C$}%
\put(452,156){\circle*{1}}%
\put(451,152){\circle*{1}}%
\put(450,148){\circle*{1}}%
\put(449,144){\circle*{1}}%
\put(448,140){\circle*{1}}%
\put(447,136){\circle*{1}}%
\put(446,132){\circle*{1}}%
\put(445,128){\circle*{1}}%
\put(444,124){\circle*{1}}%
\put(443,120){\circle*{3}}%
\put(437,105){$\cal B$}%
\end{picture}%
\caption{\small Events $\cal P$ and $\cal Q$ take place at the
same spatial point ($\cal A$) in the reference frame $(t,x)$, but
at different points ($\cal B$ and $\cal C$) in the reference frame
$(\bar{t},\bar{x})$.  This is true both in Lorentz (left) and in
Galilean (right) kinematics.}%
\label{fig1}%
\end{figure}

Energy conservation in one inertial frame, together with the
relativity principle, implies energy conservation in all inertial
frames.  As we have seen, this leads to momentum conservation.  In
fact, on replacing Eq.\ (\ref{momentum}) into Eq.\ (\ref{exp}),
one finds%
\begin{equation}%
T(\bar{u})=T(u)+v\,p(u)+{\cal O}(v^2)\;,%
\label{exp-gen}%
\end{equation}%
which holds in general.  In particular, Eq.\ (\ref{exp-gen}) is
consistent with the law of transformation for energy both in
Newtonian and Einstein dynamics, where%
\begin{equation}%
T(\bar{u})=T(u)+v\,p(u)+\frac{1}{2}\,mv^2%
\label{GalpEtransf}%
\end{equation}%
and%
\begin{equation}%
E(\bar{u})=\gamma(v)\left(E(u)+v\,p(u)\right)\;,%
\label{pEtransf}%
\end{equation}%
respectively, with $E(u)=mc^2+T(u)$ in the second case.  Now, it
is well known that energy conservation is related to invariance
under time translations, while momentum conservation is related to
invariance under space translations.  Hence, the relativity
principle has, apparently, the effect of generating invariance
under space translations from the invariance under time
translations.  (In other words, homogeneity of time in all
inertial frames enforces also homogeneity of space.)  This is
indeed the case, as one can easily understand thinking that what
appears purely as a time displacement in an inertial frame,
acquires a spatial component in any other frame with $v\neq 0$
(see Fig.\ \ref{fig1}). This is true in general, not only for a
Lorentz transformation. For example, for a Galilean transformation
between two frames $\cal K$ and $\overline{\cal K}$ one has
$\bar{x}=x+v\,t$.  Then, if two events have time and space
separations $\Delta t\neq 0$ and $\Delta x=0$ in $\cal K$, their
space separation in the reference frame $\overline{\cal K}$ is
$\Delta\bar{x}=v\,\Delta t\neq 0$, as shown in the part of Fig.\
\ref{fig1} on the right.%

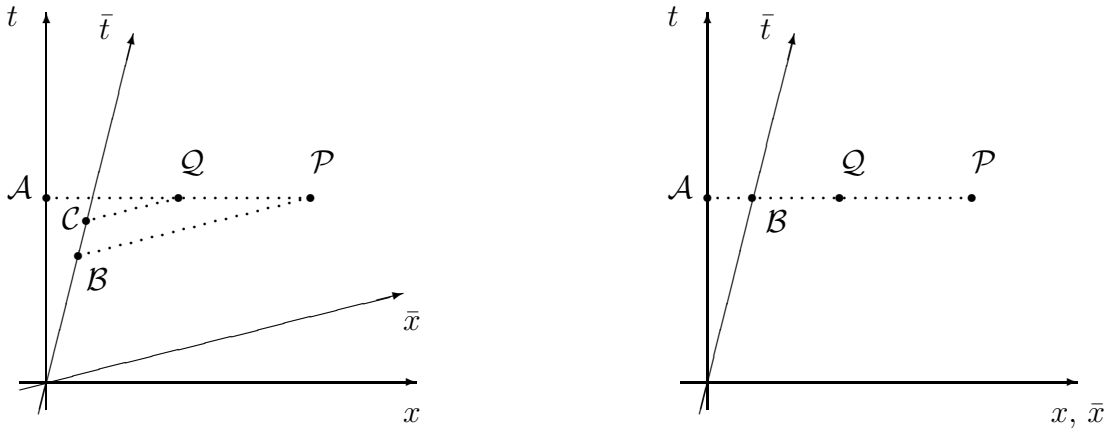
\begin{figure}%
\begin{picture}(150,200)(90,90)%
\put(110,120){\vector(1,0){150}}%
\put(255,105){$x$}%
\put(120,110){\vector(0,1){150}}%
\put(105,255){$t$}%
\put(120,120){\vector(4,1){135}}%
\put(120,120){\line(-4,-1){10}}%
\put(255,140){$\bar x$}%
\put(120,120){\vector(1,4){33}}%
\put(120,120){\line(-1,-4){3}}%
\put(140,250){$\bar t$}%
\put(170,190){\circle*{3}}%
\put(170,200){$\cal Q$}%
\put(220,190){\circle*{3}}%
\put(220,200){$\cal P$}%
\put(216,190){\circle*{1}}%
\put(212,190){\circle*{1}}%
\put(208,190){\circle*{1}}%
\put(204,190){\circle*{1}}%
\put(200,190){\circle*{1}}%
\put(196,190){\circle*{1}}%
\put(192,190){\circle*{1}}%
\put(188,190){\circle*{1}}%
\put(184,190){\circle*{1}}%
\put(180,190){\circle*{1}}%
\put(176,190){\circle*{1}}%
\put(172,190){\circle*{1}}%
\put(168,190){\circle*{1}}%
\put(164,190){\circle*{1}}%
\put(160,190){\circle*{1}}%
\put(156,190){\circle*{1}}%
\put(152,190){\circle*{1}}%
\put(148,190){\circle*{1}}%
\put(144,190){\circle*{1}}%
\put(140,190){\circle*{1}}%
\put(136,190){\circle*{1}}%
\put(132,190){\circle*{1}}%
\put(128,190){\circle*{1}}%
\put(124,190){\circle*{1}}%
\put(120,190){\circle*{3}}%
\put(105,190){$\cal A$}%
\put(216,189){\circle*{1}}%
\put(212,188){\circle*{1}}%
\put(208,187){\circle*{1}}%
\put(204,186){\circle*{1}}%
\put(200,185){\circle*{1}}%
\put(196,184){\circle*{1}}%
\put(192,183){\circle*{1}}%
\put(188,182){\circle*{1}}%
\put(184,181){\circle*{1}}%
\put(180,180){\circle*{1}}%
\put(176,179){\circle*{1}}%
\put(172,178){\circle*{1}}%
\put(168,177){\circle*{1}}%
\put(164,176){\circle*{1}}%
\put(160,175){\circle*{1}}%
\put(156,174){\circle*{1}}%
\put(152,173){\circle*{1}}%
\put(148,172){\circle*{1}}%
\put(144,171){\circle*{1}}%
\put(140,170){\circle*{1}}%
\put(136,169){\circle*{1}}%
\put(132,168){\circle*{3}}%
\put(135,155){$\cal B$}%
\put(166,189){\circle*{1}}%
\put(162,188){\circle*{1}}%
\put(158,187){\circle*{1}}%
\put(154,186){\circle*{1}}%
\put(150,185){\circle*{1}}%
\put(146,184){\circle*{1}}%
\put(142,183){\circle*{1}}%
\put(138,182){\circle*{1}}%
\put(135,181){\circle*{3}}%
\put(126,178){$\cal C$}%
\put(360,120){\vector(1,0){150}}%
\put(500,105){$x$, $\bar x$}%
\put(370,110){\vector(0,1){150}}%
\put(355,255){$t$}%
\put(370,120){\vector(1,4){33}}%
\put(370,120){\line(-1,-4){3}}%
\put(390,250){$\bar t$}%
\put(420,190){\circle*{3}}%
\put(420,200){$\cal Q$}%
\put(470,190){\circle*{3}}%
\put(470,200){$\cal P$}%
\put(466,190){\circle*{1}}%
\put(462,190){\circle*{1}}%
\put(458,190){\circle*{1}}%
\put(454,190){\circle*{1}}%
\put(450,190){\circle*{1}}%
\put(446,190){\circle*{1}}%
\put(442,190){\circle*{1}}%
\put(438,190){\circle*{1}}%
\put(434,190){\circle*{1}}%
\put(430,190){\circle*{1}}%
\put(426,190){\circle*{1}}%
\put(422,190){\circle*{1}}%
\put(418,190){\circle*{1}}%
\put(414,190){\circle*{1}}%
\put(410,190){\circle*{1}}%
\put(406,190){\circle*{1}}%
\put(402,190){\circle*{1}}%
\put(398,190){\circle*{1}}%
\put(394,190){\circle*{1}}%
\put(390,190){\circle*{1}}%
\put(387,190){\circle*{3}}%
\put(392,177){$\cal B$}%
\put(382,190){\circle*{1}}%
\put(378,190){\circle*{1}}%
\put(374,190){\circle*{1}}%
\put(370,190){\circle*{3}}%
\put(355,190){$\cal A$}%
\end{picture}%
\caption{\small Events $\cal P$ and $\cal Q$ take place at the
same time (point $\cal A$ in the diagrams) according to the
reference frame $(t,x)$.  In a different frame $(\bar{t},\bar{x})$
they occur at different times in Einstein kinematics (points $\cal
B$ and $\cal C$ in the diagram on the left), while they are still
simultaneous if one uses Galilean kinematics (point $\cal B$ in
the diagram on the right).}%
\label{fig2}%
\end{figure}

The situation is not symmetric regarding momentum conservation, as
one can see already by examining the cases of a Lorentz and a
Galilean transformation.  In the first case,%
\begin{equation}%
p(\bar{u})=\gamma(v)\left(p(u)+v\,E(u)/c^2\right)%
\label{pLorentz}%
\end{equation}%
implies%
\begin{equation}%
\frac{\d p(u)}{\d u}\,\varphi(u)=\frac{E(u)}{c^2}\;.%
\label{derivatives}%
\end{equation}%
Hence, one can enforce energy conservation by requiring momentum
conservation in every inertial frame.\footnote{Interestingly, one
can combine Eqs.\ (\ref{momentum}) and (\ref{derivatives}) to get
a single differential equation for $p$,%
\[\frac{\d^2 p}{\d U^2}-\frac{1}{c^2}\,p=0\;,\]%
where $U=h(u)$, with $h$ given by Eq.\ (\ref{U}) with $K=1/c^2$.}
On the other hand, in Newtonian mechanics Eq.\ (\ref{pLorentz}) is
replaced by%
\begin{equation}%
p(\bar{u})=p(u)+m\,v\;,%
\label{pGal}%
\end{equation}%
from which one gets%
\begin{equation}%
\frac{\d p(u)}{\d u}\,\varphi(u)=m\;.\\%
\label{Galderivatives}%
\end{equation}%
Therefore, momentum conservation in all inertial frames now
enforces conservation of mass, rather than of energy.  This
asymmetry is related to the fact that, while under a Lorentz
transformation a purely spatial displacement acquires also a time
component, this is not true for a Galilean transformation (see
Fig.\ \ref{fig2}). Indeed, since for the latter one has
$\bar{t}=t$, it will be $\Delta\bar{t}=\Delta t$ regardless
of what $\Delta x$ is.%

Of course, there is no reason to stop the analysis in Sec.\
\ref{subsec:pT} to the first order in $v$.  In fact, by
considering the second order, then the third order, and so on, an
infinite set of conserved quantities can be generated. However,
these ``extra'' quantities are not independent and do not give
anything new, as one might also expect noticing that conservation
of energy and momentum already exploit all the available
symmetries, namely, homogeneity of time and space.  It is
nevertheless instructive to see explicitly what happens for the
two dynamics considered in Sec.\ \ref{sec:examples}.%

In Newtonian dynamics, at the second order in $v$ one recovers
conservation of mass (which is already contained, however, in our
implicit assumption that masses do not change during a collision),
while at still higher orders all coefficients vanish identically
--- see Eq.\ (\ref{T'newt}). In Einstein dynamics, at the second
order one finds conservation of energy $E$ which, once again,
amounts to conservation of mass when one considers that $T$ is
also conserved, by postulate P2. At orders higher than two the
situation is a bit more involved. It is convenient first to
rewrite the second equation in (\ref{pEtransf}) as%
\begin{equation}%
E(\Phi(u,v))=E(u)\,\gamma(v)+p(u)\,\gamma(v)\,v\;.%
\label{E(Phi)}%
\end{equation}%
Now, the coefficient $\gamma(v)$ which appears on the right-hand
side of Eq.\ (\ref{E(Phi)}) is a function of $v^2$, so the
coefficients of the expansion of $E(\Phi(u,v))$ in powers of $v$
will all be equal to $E(u)$ for even powers, and to $p(u)$ for odd
powers.  Since $E$ and $T$ differ only by a velocity-independent
quantity, one simply recovers the conservation of energy and
momentum, alternatively.%

\section*{Acknowledgements}

We thank an anonymous board member for stimulating several
improvements in the presentation.  SS is grateful to Daniela
W\"unsch for bringing Ref.\ \cite{kaluza} to his knowledge.%

%
\end{document}